# On the Shapley-like Payoff Mechanisms in Peer-Assisted Services with Multiple Content Providers

Jeong-woo Cho and Yung Yi

Dept. of Electrical Engineering, KAIST, South Korea

{ggumdol,yiyung}@kaist.ac.kr

**Abstract**

This paper studies an incentive structure for cooperation and its stability in peer-assisted services when there exist multiple content providers, using a coalition game theoretic approach. We first consider a generalized coalition structure consisting of multiple providers with many assisting peers, where peers assist providers to reduce the operational cost in content distribution. To distribute the profit from cost reduction to players (*i.e.*, providers and peers), we then establish a generalized formula for individual payoffs when a "Shapley-like" payoff mechanism is adopted. We show that the grand coalition is *unstable*, even when the operational cost functions are concave, which is in sharp contrast to the recently studied case of a single provider where the grand coalition is stable. We also show that irrespective of stability of the grand coalition, there always exist coalition structures which are not convergent to the grand coalition. Our results give us an important insight that a provider does not tend to cooperate with other providers in peer-assisted services, and be separated from them. To further study the case of the separated providers, three examples are presented; *(i)* underpaid peers, *(ii)* service monopoly, and *(iii)* oscillatory coalition structure. Our study opens many new questions such as realistic and efficient incentive structures and the tradeoffs between fairness and individual providers' competition in peer-assisted services.

## I. Introduction

The Internet is becoming more content-oriented, and the need of cost-effective and scalable distribution of contents has become the central role of the Internet. Uncoordinated peer-to-peer (P2P) systems, *e.g.*, BitTorrent, has been successful in distributing contents, but the rights of the content owners are not protected well, and most of the P2P contents are in fact illegal. In its response, a new type of service, called *peer-assisted services*, has received significant attentions these days. In peer-assisted services, users commit a part of their resources to assist content providers in content distribution with objective of enjoying both scalability/efficiency in P2P systems and controllability in client-server systems. Examples of peer-assisted services include nano data center [1] and IPTV [2], where high potential of operational cost reduction was observed. For instance, there are now 1.8 million IPTV subscribers in South Korea, and the financial sectors forecast that by 2014 the IPTV subscribers is expected to be 106 million, see, *e.g.*, [3]. However, it is clear that most users will not just "donate" their resources to content providers. Thus, the key factor to the success of peer-assisted services is how to (economically) incentivize users to commit their valuable resources and participate in the service.

One of nice mathematical tools to study incentive-compatibility of peer-assisted services is the coalition game theory which covers how payoffs should be distributed and whether such a payoff scheme can be executed by rational individuals or not. In peer-assisted services, the "symbiosis" between providers and peers are sustained when *(i)* the offered payoff scheme guarantees fair assessment of players' contribution under a provider-peer coalition and *(ii)* each individual has no incentive to exit from the coalition. In the coalition game theory, the notions of Shapley value and the core have been popularly applied to address *(i)* and *(ii)*, respectively, when the entire players cooperate, referred to as the grand coalition. A recent paper by Misra *et al.* [4] demonstrates that the Shapley value approach is a promising payoff mechanism to provide right incentives for cooperation in a *single-provider* peer-assisted service.

However, in practice, the Internet consists of multiple content providers, even if only giant providers are counted. The focus of our paper is to study the cooperation incentives for *multiple* providers. In the multi-provider case, the model clearly becomes more complex, thus even classical analysis adopted in the single-provider case becomes much more challenging, and moreover the results and their implications may experience drastic changes. To motivate further, see an example in Fig. 1 with two providers (Google TV and iTunes) and consider two cases of cooperation: *(i) separated,* where there exists a fixed partition of peers for each provider, and *(ii) coalescent,* where each peer is possible to assist any provider. In the separated case, a candidate payoff scheme is based on the Shapley value in each separated coalition. Similarly, in the coalescent case, the Shapley value is also a candidate payoff scheme after the worth function of the grand coalition $N$ (the player set) is defined

This work was supported by Brain Korea 21 Project, BK Electronics and Communications Technology Division, KAIST in 2011, and KRCF (Korea Research Council of Fundamental Science and Research).

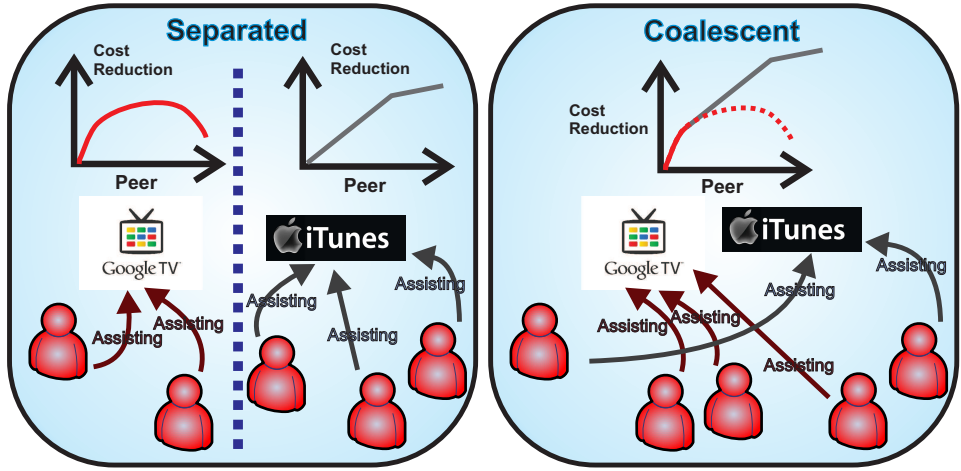

Fig. 1. Coalition Structures for a Dual-Provider Network.

appropriately. A reasonable definition of the worth function[1] can be the total cost reduction, maximized over all combinations of peer partitions to each provider[2]. Then, it is not hard to see that the cost reduction for the coalescent case exceeds that for the separated case, unless the two partitions are equivalent in both cases. This implies that at least one individual in the separated case is *underpaid* than in the coalescent case under the Shapley-value based payoff mechanism. Thus, providers and users are recommended to form the grand coalition and be paid off based on the Shapley value, *i.e.*, the due desert.

However, it is still questionable whether peers will stay in the grand coalition and thus the consequent Shapley-value based payoff mechanism is desirable in the multi-provider setting. In this paper, we anatomize incentive structures in peer-assisted services with multiple content providers and focus on stability issues from two different angles: stability at equilibrium of Shapley value and convergence to the equilibrium.

Our main contributions are summarized as follows:

1) We first provide a closed-form formula of the Shapley value for a general case of multiple providers and peers. To that end, we define a worth function to be the maximum total cost reduction over all possible peer partitions to each provider. Due to the intractability of analytical computation of the Shapley value, we take a fluid-limit approximation that assumes a large number of peers and re-scales the system with the number of peers. This is a non-trivial generalization of the Shapley value for the single-provider case in [4]. In fact, our formula in Theorem 1 establishes the general Shapley value for distinguished *multiple* atomic players and infinitesimal players in the context of the Aumann-Shapley (A-S) prices [5] in coalition game theory.

2) We prove in Theorem 2 that the Shapley value for the multiple-provider case is not in the core under mild conditions, *e.g.*, each provider's cost function is concave. This is in stark contrast to the single-provider case where the concave cost function stabilizes the equilibrium.

3) We study, for the first time, the endogenous formation of coalitions in peer-assisted services by introducing the stability notion defined by the seminal work of Hart and Kurz [6]. We show that, if we adopt a Shapley-like payoff mechanism, called Aumann-Drèze value, irrespective of stability of the grand coalition, there always exist initial states which are not convergent to the grand coalition. An interesting fact from this part of study is that peers and providers have opposite cooperative preferences, *i.e.*, peers prefer to cooperate with more providers, whereas providers prefer to be separated from other providers.

In short, the Shapley payoff regime cannot incentivize rational players to form the grand coalition, implying that *fair* profit-sharing and *opportunism* of players cannot stand together. If the grand coalition is broken up, the Shapley payoff scheme cannot be executed because the total profit of a coalition *differs* from the sum of Shapley payoffs in the coalition. Only payoff mechanisms for general coalition structure may be used. In conjunction with this point, we present three examples for non-cooperation among providers who adopt the Shapley-like payoff scheme: *(i)* the peers are underpaid than their Shapley payoffs, *(ii)* a provider with more "advantageous" cost function monopolizes all peers, and *(iii)* Shapley value for each coalition gives rise to an oscillatory behavior of coalition structures. These examples suggest that the system with the separated providers may be even unstable as well as unfairness in a peer-assisted service market.

The rest of the paper is organized as follows. In Section II, we define Shapley and Aumann-Drèze values with minimal formalism. After formulating the fluid Aumann-Drèze formula for multiple providers, we establish results on the stability-related concepts in Section IV to substantiate that it is very unlikely that the grand coalition occurs. Then we point out main drawbacks of Aumann-Drèze value in Section V and conclude this paper.

---

[1]We establish in Section III-A that this definition is *derived* directly from an essential property of coalition.

[2]The notion of peer partitions implicitly assumes that a peer assists only one provider. However, our model is not restricted in the sense that we will study the regime of a large number of peers for mathematical tractability, in which case a peer assisting two providers can be regarded as two distinct peers assisting distinct providers.

## II. PRELIMINARIES

Since this paper investigates a multi-provider case, where a peer can choose any provider to assist, we start this section by defining a coalition game with a peer partition (*i.e.*, a coalition structure) and introducing the payoff mechanism thereof.

### A. Game with Coalition Structure

A game with coalition structure is a triple $(N, v, \mathcal{P})$ where $N$ is a player set and $v : 2^N \to \mathbb{R}$ ($2^N$ is the set of all subsets of $N$) is a worth function, $v(\emptyset) = 0$. $v(K)$ is called the worth of a coalition $K \subseteq N$. $\mathcal{P}$ is called a *coalition structure* for $(N, v)$; it is a partition of $N$ where $C(i) \in \mathcal{P}$ denotes the coalition containing player $i$. For your reference, a coalition structure $\mathcal{P}$ can be regarded as a set of disjoint coalitions. The *grand coalition* is the partition $\mathcal{P} = \{N\}$. For instance[3], a partition of $N = \{1, 2, 3, 4, 5\}$ is $\mathcal{P} = \{\{1, 2\}, \{3, 4, 5\}\}$, $C(4) = \{3, 4, 5\}$, and the grand coalition is $\mathcal{P} = \{\{1, 2, 3, 4, 5\}\}$. $\mathcal{P}(N)$ is the set of all partitions of $N$. For notational simplicity, a game *without* coalition structure $(N, v, \{N\})$ is denoted by $(N, v)$. A value of player $i$ is an operator $\phi_i(N, v, \mathcal{P})$ that assigns a payoff to player $i$. We define $\phi_K = \sum_{i \in K} \phi_i$ for all $K \subseteq N$.

To conduct the equilibrium analysis of coalition games, the notion of *core* has been extensively used to study the stability of grand coalition $\mathcal{P} = \{N\}$:

**Definition 1 (Core)** The core is defined by $\{\phi(N, v) \mid \sum_{i \in N} \phi_i(N, v) = v(N) \text{ and } \sum_{i \in K} \phi_i(N, v) \geq v(K), \forall K \subseteq N\}$.

If a payoff vector $\phi(N, v)$ lies in the core, no player in $N$ has an incentive to split off to form another coalition $K$ because the worth of the coalition $K$, $v(K)$, is no more than the payoff sum $\sum_{i \in K} \phi_i(N, v)$. Note that the definition of the core hypothesizes that the grand coalition is already formed *ex-ante*. We can see the core as an analog of Nash equilibrium from noncooperative games. Precisely speaking, it should be viewed as an analog of *strong Nash equilibrium* where no arbitrary coalition of players can create worth which is larger than what they receive in the grand coalition. If a payoff vector $\phi(N, v)$ lies in the core, then the grand coalition is stable with respect to any collusion to break the grand coalition.

### B. Shapley Value and Aumann-Drèze Value

On the premise that the player set is not partitioned, *i.e.*, $\mathcal{P} = \{N\}$, the Shapley value, denoted by $\varphi$ (not $\phi$), is popularly used as a fair distribution of the grand coalition's worth to individual players, defined by:

$$\varphi_i(N, v) = \sum_{S \subseteq N \setminus \{i\}} \frac{|S|!(|N| - |S| - 1)!}{|N|!} \left(v(S \cup \{i\}) - v(S)\right). \tag{1}$$

Shapley [7] gives the following interpretation: "*(i)* Starting with a single member, the coalition adds one player at a time until everybody has been admitted. *(ii)* The order in which players are to join is determined by chance, with all arrangements equally probable. *(iii)* Each player, on his admission, demands and is promised the amount which his adherence contributes to the value of the coalition." The Shapley value quantifies the above that is axiomatized (see Section II-C) and has been treated as a worth distribution scheme. The beauty of the Shapley value lies in that the payoff "summarizes" in *one* number all the possibilities of each player's contribution in every coalition structure.

Given a coalition structure $\mathcal{P} \neq \{N\}$, one can obtain the Aumann-Drèze value (A-D value) [8] of player $i$, also denoted by $\varphi$, by taking $C(i)$, which is the coalition containing player $i$, to be the player set and by computing the Shapley value of player $i$ of the *reduced* game $(C(i), v)$. It is easy to see that the A-D value can be construed as a direct extension of the Shapley value to a game with coalition structure. Note that both Shapley value and A-D value are denoted by $\varphi$ because the only difference is the underlying coalition structure $\mathcal{P}$.

### C. Axiomatic Characterizations of Values

We provide here the original version [7] of the axiomatic characterization of the Shapley value.

**Axiom 1 (Coalition Efficiency, CE)** For all $C \in \mathcal{P}$, $\sum_{j \in C} \phi_j(N, v, \mathcal{P}) = v(C)$.
**Axiom 2 (Coalition Restricted Symmetry, CS)** If $j \in C(i)$ and $v(K \cup \{i\}) = v(K \cup \{j\})$ for all $K \subseteq N \setminus \{i, j\}$, then $\phi_i(N, v, \mathcal{P}) = \phi_j(N, v, \mathcal{P})$.
**Axiom 3 (Additivity, ADD)** $\phi_i(N, v + v', \mathcal{P}) = \phi_i(N, v, \mathcal{P}) + \phi_i(N, v', \mathcal{P})$ for all coalition functions $v$, $v'$ and $i \in N$.
**Axiom 4 (Null Player, NP)** If $v(K \cup \{i\}) = v(K)$ for all $K \subseteq N$, then $\phi_i(N, v, \mathcal{P}) = 0$.

Recall that the basic premise of the Shapley value is that the player set is not partitioned, *i.e.*, $\mathcal{P} = \{N\}$. Also, the Shapley value, defined in (1), is *uniquely* characterized by **CE**, **CS**, **ADD** and **NP** for $\mathcal{P} = \{N\}$ [7]. The A-D value is also *uniquely* characterized by **CE**, **CS**, **ADD** and **NP** (Axioms 1-4), but in this case for arbitrary coalition structure $\mathcal{P}$ [8]. In the literature, *e.g.*, [9], [10], the A-D value has been used to analyze the *static* games where a coalition structure is *exogenously* given.

---
[3] A player $i$ is an *element* of a coalition $C = C(i)$, which is in turn an *element* of a partition $\mathcal{P}$. $\mathcal{P}$ is an element of $\mathcal{P}(N)$ *while* a subset of $2^N$.

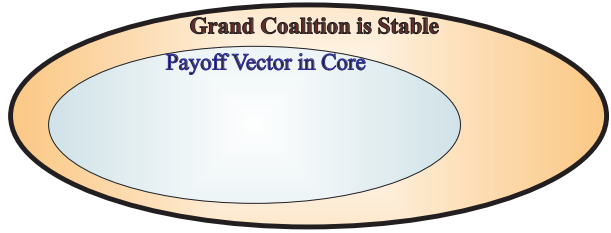

Fig. 2. If a payoff vector lies in the core, the grand coalition is stable [6].

**Definition 2 (Coalition Independent, CI)** If $i \in C \subseteq N$, $C \in \mathcal{P}$ and $C \in \mathcal{P}'$, then $\phi_i(N, v, \mathcal{P}) = \phi_i(N, v, \mathcal{P}')$.

From the definition of the A-D value, the payoff of player $i$ in coalition $C(i)$ is affected neither by the player set $N$ nor by coalitions $C \in \mathcal{P}$, $C \neq C(i)$. Note that only $C(i)$ contains the player $i$. Thus, it is easy to prove that the A-D value is coalition independent. From **CI** of the A-D value, in order to decide the payoffs of a game with general coalition structure $\mathcal{P}$, it suffices to decide the payoffs of players within each coalition, say $C \in \mathcal{P}$, without considering other coalitions $C \in \mathcal{P}$, $C \neq C(i)$. In other words, once we decide the payoffs of a coalition $C \in \mathcal{P}$, the payoffs remains unchanged even though other coalitions, $C' \in \mathcal{P}$, $C' \neq C$, vary. Thus, for any given coalition structure $\mathcal{P}$, any coalition $C \in \mathcal{P}$ is just two-fold in terms of the number of providers in $C$: *(i)* one provider or *(ii)* two or more providers, as depicted in Fig. 1.

The notion of **CI** also enables us to define the stability of a game with coalition structure in the following simplistic way:

**Definition 3 (Stable Coalition Structure [6])** We say that a coalition structure $\mathcal{P}'$ blocks $\mathcal{P}$, where $\mathcal{P}'$, $\mathcal{P} \in \mathcal{P}(N)$, with respect to $\phi$ if and only if there exists some $C \in \mathcal{P}'$ such that $\phi_i(N, v, \{C, \cdots\}) > \phi_i(N, v, \mathcal{P})$ for all $i \in C$. In this case, we also say that $C$ blocks $\mathcal{P}$. If there does not exist any $\mathcal{P}'$ which blocks $\mathcal{P}$, $\mathcal{P}$ is called *stable*.

Due to **CI** of the A-D value, all stability notions defined by the seminal work of Hart and Kurz [6] coincide with the above simplistic definition, as discussed by Tutic [11]. Definition 3 can be intuitively interpreted that, if there exists any subset of players $C$ who improve their payoffs away from the current coalition structure, they *will* form a new coalition $C$. In other words, if a coalition structure $\mathcal{P}$ has any blocking coalition $C$, some rational players will break $\mathcal{P}$ to increase their payoffs. The basic premise here is that players are not clairvoyant, *i.e.*, they are interested only in improving their instant payoffs. If a payoff vector lies in the core, the grand coalition is stable in the sense of Definition 3, but the converse is not necessarily true (see Fig. 2).

## III. COALITION GAME IN PEER-ASSISTED SERVICES

In this section, we first define a coalition game in a peer-assisted service with multiple content providers by classifying the types of coalition structures as *separated*, where a coalition includes only one provider, and *coalescent*, where a coalition is allowed to include more than one providers (see Fig. 1). To define the coalition game, we will define a worth function of an arbitrary coalition $S \subseteq N$ for such two cases.

### A. Worth Function in Peer-Assisted Services

Assume that players $N$ are divided into two sets, the set of content providers $Z := \{p_1, \cdots, p_\zeta\}$, and the set of peers $H := \{n_1, \cdots, n_\eta\}$, *i.e.*, $N = Z \cup H$. We also assume that the peers are homogeneous, *e.g.*, the same computing powers, disk cache sizes, and upload bandwidths. Later, we discuss that our results can be readily extended to nonhomogeneous peers. The set of peers assisting providers is denoted by $\bar{H} := \{n_1, \cdots, n_{x \cdot \eta}\}$ where $x = |\bar{H}|/\eta$, *i.e.*, the fraction of assisting peers. We define the worth of a coalition $S$ to be the amount of cost reduction due to cooperative distribution of the contents by the players in $S$ in both separated and coalescent cases.

**Separated case**: Denote by $\Omega_p^\eta(x(S))$ the operational cost of a provider $p$ when the coalition $S$ consists of provider $p$ and $x(S) \cdot \eta$ assisting peers. Since the operational cost cannot be negative, we assume $\Omega_p^\eta(x(S)) > 0$. Note that from the homogeneity assumption of peers, the cost function depends only on the fraction of assisting peers. Then, we define the worth function $\hat{v}(S)$ for the coalition $S$ as:

$$\hat{v}(S) := \Omega_p^\eta(0) - \Omega_p^\eta(x(S)) \qquad (2)$$

where $\Omega_p^\eta(0)$ corresponds to the cost when there are no assisting peers. For notational simplicity, in what follows, $x(S)$ is denoted by $x$ from now on.

**Coalescent case**: In contrast to the separated case, where a coalition includes a single provider, the worth for the coalescent case is not clear yet, since depending on which peers assist which providers the amount of cost reduction may differ. One of reasonable definitions would be the maximum worth out of all peer partitions, *i.e.*, the worth for the coalescent case is defined by:

$$v(S) = \max \left\{ \sum_{C \in \mathcal{P}} \hat{v}(C) \ \Big| \ \mathcal{P} \in \mathcal{P}(S) \text{ such that } |Z \cap C| = 1, \ \forall C \in \mathcal{P} \right\}. \tag{3}$$

The definition above implies that we *view* a coalition containing more than one provider as the most productive coalition whose worth is *maximized* by choosing the optimal partition $\mathcal{P}^*$ among all possible partitions of $S$. Note that (3) is consistent with the definition (2) for $|Z \cap S| \leq 1$, *i.e.*, $v(S) = \hat{v}(S)$ for $|Z \cap S| \leq 1$.

Four remarks are in order. First, as opposed to [4] where $\hat{v}(\{p\}) = \eta R - \Omega_p^\eta(0)$ ($R$ is the subscription fee paid by any peer), we simply assume that $\hat{v}(\{p\}) = 0$. Note that, as discussed in [10, Chapter 2.2.1], it is no loss of generality to assume that, initially, each provider has earned no money. In our context, this means that it does not matter how much fraction of peers is subscribing to each provider because each peer has already paid the subscription fee to providers *ex-ante*.

Second, it is also important to note that we cannot always assume that $\Omega_p^\eta(x)$ is monotonically decreasing because providers have to pay the electricity expense of the computers and the maintenance cost of the hard disks of assisting peers. For example, a recent study [12] found that Annualized Failure Rate (AFR) of hard disk drives is over 8.6% for three-year old ones. We discuss in Appendix A that, if we consider a more *intelligent* coalition, the cost function $\Omega_p^\eta(x)$ is always non-increasing. However, we assume the following to simplify the exposition:

**Assumption 1** $\Omega_p^\eta(x)$ is non-increasing in $x$ for all $p \in Z$.

Third, the worth function in peer-assisted services can reflect the diversity of peers. It is not difficult to extend our result to the case where peers belong to distinct classes. For example, peers may be distinguished by different upload bandwidths and different hard disk cache sizes. A point at issue for the multiple provider case is whether peers who are *not* subscribing to the content of a provider may be allowed to assist the provider or not. On the assumption that the content is ciphered and not decipherable by the peers who do not know its password which is given only to the subscribers, providers will allow those peers to assist the content distribution. Otherwise, we can easily reflect this issue by dividing the peers into a number of classes where each class is a set of peers subscribing to a certain content.

Lastly, it should be pointed out that the worth function in (3) is *rigorously* selected in order to satisfy a basic property:

**Definition 4 (Superadditivity)** A worth function $v$ is superadditive if $(S, T \subseteq N \text{ and } S \cap T = \emptyset) \Rightarrow v(S \cup T) \geq v(S) + v(T)$.

Suppose we have a superadditive worth function $v'$. We require $v'(S) = \hat{v}(S)$ if $S$ includes one provider. It follows from the definition of $v$ in (3) that $v'(\cdot)$ is no greater than $v(\cdot)$, *i.e.*, $v(\cdot) \geq v'(\cdot)$ because $v$ is the total cost reduction that is maximized over all possible peer partitions to each provider. In the meantime, since $v'$ is superadditive, it must satisfy $v'(S \cup T) \geq v'(S) + v'(T)$ for all disjoint $S, T \subseteq N$, implying that $v'(\cdot) \geq v(\cdot)$. This completes the proof of the following lemma.

**Lemma 1** When the worth is given by (2), there exists a superadditive worth function, uniquely given by (3).

Superadditivity is one of the most elementary properties, which ensures that the core is nonempty by appealing to Bondareva-Shapley Theorem [10, Theorem 3.1.4]. In light of this lemma, we can clearly restate that our objective in this paper is to analyze the incentive structure of peer-assisted services when the worth of coalition is superadditive. This objective then necessarily implies the form of worth function in (3).

### B. Fluid Aumann-Drèze Value for Multiple-Provider Coalitions

So far we have defined the worth of coalitions. Now let us *distribute* the worth to the players for a given coalition structure $\mathcal{P}$. Recall that the payoffs of players in a coalition are independent from other coalitions by the definition of A-D payoff. Pick a coalition $C$ without loss of generality, and denote the set of providers in $C$ by $\bar{Z} \subseteq Z$. With slight notational abuse, the set of peers assisting $\bar{Z}$ is denoted by $\bar{H}$. Once we find the A-D payoff for a coalition consisting of arbitrary provider set $\bar{Z} \subseteq Z$ and assisting peer set $\bar{H} \subseteq H$, the payoffs for the separated and coalescent cases in Fig. 1 follow from the substitutions $\bar{Z} = Z$ and $\bar{Z} = \{p\}$, respectively. In light of our discussion in Section II-B, it is more reasonable to call a Shapley-like payoff mechanism 'A-D payoff' and 'Shapley payoff' respectively for the partitioned and non-partitioned games $(N, v, \{\bar{Z} \cup \bar{H}, \cdots\})$ and $(N, v, \{Z \cup H\})$[4].

**Fluid Limit**: We adopt the limit axioms for a large population of users to overcome the computational hardness of the A-D payoffs:

$$\lim_{\eta \to \infty} \widetilde{\Omega}_p^\eta(\cdot) = \widetilde{\Omega}_p(\cdot) \text{ where } \widetilde{\Omega}_p^\eta(\cdot) = \tfrac{1}{\eta} \Omega_p^\eta(\cdot) \tag{4}$$

which is the asymptotic operational cost per peer in the system with a large number of peers. We drop superscript $\eta$ from notations to denote their limits as $\eta \to \infty$. From the assumption $\Omega_p^\eta(x) > 0$, we have $\widetilde{\Omega}_p(x) \geq 0$. To avoid trivial cases, we also assume $\widetilde{\Omega}_p(x)$ is not constant in the interval $x \in [0, 1]$ for any $p \in Z$. We also introduce the payoff of each provider per user, defined as $\widetilde{\varphi}_p^\eta := \frac{1}{\eta}\varphi_p^\eta$. We now derive the fluid limit equations of the payoffs which can be obtained as $\eta \to \infty$. The proof of the following theorem is given in Appendix B.

---
[4]On the contrary, the term 'Shapley payoff' was used in [4] to refer to the payoff for the game $(N, v, \{\bar{Z} \cup \bar{H}, \cdots\})$ where a proper subset of the peer set assists the content distribution.

**Theorem 1 (A-D Payoff for Multiple Providers)** As $\eta$ tends to $\infty$, the A-D payoffs of providers and peers under an arbitrary coalition $C = \bar{Z} \cup \bar{H}$ converge to the following equation:

$$\begin{cases} \widetilde{\varphi}_p^{\bar{Z}}(x) = \widetilde{\Omega}_p(0) - \sum_{S \subseteq \bar{Z}\setminus\{p\}} \int_0^1 u^{|S|}(1-u)^{|\bar{Z}|-1-|S|} \left( M_\Omega^{S\cup\{p\}}(ux) - M_\Omega^S(ux) \right) du, & \text{for } p \in \bar{Z} \\ \widetilde{\varphi}_n^{\bar{Z}}(x) = -\sum_{S \subseteq \bar{Z}} \int_0^1 u^{|S|}(1-u)^{|\bar{Z}|-|S|} \frac{dM_\Omega^S}{dx}(ux) du, & \text{for } n \in \bar{H}. \end{cases} \quad (5)$$

Here $M_\Omega^S(x) := \min\left\{ \sum_{i \in S} \widetilde{\Omega}_i(y_i) \mid \sum_{i \in S} y_i \leq x, \ y_i \geq 0 \right\}$ and $M_\Omega^\emptyset(x) := 0$. Note that $M_\Omega^{\{p\}}(x) = \widetilde{\Omega}_p(x)$.

The following corollaries are immediate as special cases of Theorem 1, which we will use in Section V.

**Corollary 1 (A-D Payoff for Single Provider)** As $\eta$ tends to $\infty$, the A-D payoffs of providers and peers who belong to a single-provider coalition, i.e., $\bar{Z} = \{p\}$, converge to the following equation:

$$\begin{cases} \widetilde{\varphi}_p^{\{p\}}(x) = \widetilde{\Omega}_p(0) - \int_0^1 M_\Omega^{\{p\}}(ux) du, \\ \widetilde{\varphi}_n^{\{p\}}(x) = -\int_0^1 u \frac{dM_\Omega^{\{p\}}}{dx}(ux) du, & \text{for } n \in \bar{H}. \end{cases} \quad (6)$$

**Corollary 2 (A-D Payoff for Dual Providers)** As $\eta$ tends to $\infty$, the A-D payoffs of providers and peers who belong to a dual-provider coalition, i.e., $\bar{Z} = \{p, q\}$, converge to the following equation:

$$\begin{cases} \widetilde{\varphi}_p^{\{p,q\}}(x) = \widetilde{\Omega}_p(0) - \int_0^1 u M_\Omega^{\{p,q\}}(ux) du - \int_0^1 (1-u) M_\Omega^{\{p\}}(ux) du + \int_0^1 u M_\Omega^{\{q\}}(ux) du, & (p, q \text{ are interchangeable}) \\ \widetilde{\varphi}_n^{\{p,q\}}(x) = -\int_0^1 u^2 \frac{dM_\Omega^{\{p,q\}}}{dx}(ux) du - \sum_{i \in \{p,q\}} \int_0^1 u(1-u) \frac{dM_\Omega^{\{i\}}}{dx}(ux) du, & \text{for } n \in \bar{H}. \end{cases} \quad (7)$$

Note that our A-D payoff formula in Theorem 1 generalizes the formula in Misra *et al.* [4, Theorem 4.3] (*i.e.*, $|Z| = 1$). It also establishes the A-D values for distinguished *multiple* atomic players (the providers) and infinitesimal players (the peers), in the context of the Aumann-Shapley (A-S) prices [5] in coalition game theory.

Our formula for the peers can be interpreted as follows. Take the second line of (7) as an example. Recall the definition of the Shapley value (1). The payoff of peer $n$ is the *marginal* cost reduction $v(S \cup \{n\}) - v(S)$ that is *averaged* over all equally probable arrangements, *i.e.*, the orders of players. It is also implied by (1) that the *expectation* of the marginal cost is computed under the assumption that the events $|S| = y$ and $|S| = y'$ for $y \neq y'$ are *equally probable*, *i.e.*, $\mathsf{P}(|S| = y) = \mathsf{P}(|S| = y')$. Therefore, in our context of infinite player game in Theorem 1, for every values of $ux$ along the interval $[0, x]$, the subset $S \subseteq \bar{Z} \cup \bar{H}$ contains $ux$ fraction of the peers. More importantly, the probability that each provider is a member of $S$ is simply $u$ because the size of peers in $S$, $\eta ux$, is infinite as $\eta \to \infty$ so that the *size* of $S$ is not affected by whether a provider belongs to $S$ or not. Therefore, the marginal cost reduction of each peer on the condition that both providers are contained in $S$ becomes $-u^2 \frac{dM_\Omega^{\{p,q\}}}{dx}(ux)$. Likewise, the marginal cost reduction of each peer on the condition that only one provider is in the coalition is $-u(1-u) \frac{dM_\Omega^{\{p\}}}{dx}(ux)$.

## IV. INSTABILITY OF THE GRAND COALITION

In this section, we study the stability of the grand coalition to see if *rational* players are willing to form the grand coalition, only under which they can be paid their respective *fair* Shapley payoffs. The key message of this section is that the rational behavior of the providers makes the Shapley value approach *unworkable* because the major premise of the Shapley value, the grand coalition, is not formed in the multi-provider games.

### A. Stability of the Grand Coalition

Guaranteeing the stability of a payoff vector has been an important topic in coalition game theory. For the single-provider case, $|Z| = 1$, it was shown in [4, Theorem 4.2] that, if the cost function is decreasing and concave, the Shapley incentive structure lies in the core of the game. What if for $|Z| \geq 2$? Is the grand coalition stable for the multi-provider case? Prior to addressing this question, we first define the following:

**Definition 5 (Noncontributing Provider)** A provider $p \in Z$ is called *noncontributing* if $M_\Omega^Z(1) - M_\Omega^{Z\setminus\{p\}}(1) = \widetilde{\Omega}_p(0)$.

To understand this better, note that the above expression is equivalent to the following:

$$\left( \sum_{i \in Z} \widetilde{\Omega}_i(0) - M_\Omega^Z(1) \right) - \left( \sum_{i \in Z\setminus\{p\}} \widetilde{\Omega}_i(0) - M_\Omega^{Z\setminus\{p\}}(1) \right) = 0 \quad (8)$$

which implies that there is no difference in the total cost reduction, irrespective of whether the provider $p$ is in the provider set or not. Interestingly, if all cost functions are concave, there exists at least one noncontributing provider.

**Lemma 2** Suppose $|Z| \geq 2$. If $\widetilde{\Omega}_p(\cdot)$ is concave for all $p \in Z$, there exist $|Z| - 1$ noncontributing providers.

To prove this, recall the definition of $M_\Omega^Z(\cdot)$:

$$M_\Omega^Z(x) = \min_{y \in Y(x)} \sum_{i \in Z} \widetilde{\Omega}_i(y_i) \quad \text{where } Y(x) := \{(y_1, \cdots, y_{|Z|}) \mid \sum_{i \in Z} y_i \leq x, \ y_i \geq 0\}.$$

Since the summation of concave functions is concave and the minimum of a concave function over a convex feasible region $Y(x)$ is an *extreme* point of $Y(x)$ as shown in [13, Theorem 3.4.7], we can see that the solutions of the above minimization are the extreme points of $\{(y_1, \cdots, y_{|Z|}) \mid \sum_{i \in Z} y_i \leq x, \ y_i \geq 0\}$, which in turn imply $y_i = 0$ for $|Z| - 1$ providers in $Z$. Note that the condition $|Z| \geq 2$ is *necessary* here.

We are ready to state the following theorem, a direct consequence of Theorem 1. The proof is given in Appendix C.

**Theorem 2 (Shapley Payoff Not in the Core)** If there exists a noncontributing provider, the Shapley payoff for the game $(Z \cup H, v)$ does not lie in the core.

It follows from Lemma 2 that, if all operational cost functions are concave and $|Z| \geq 2$, the Shapley payoff does not lie in the core. This result appears to be in best agreement with our usual intuition. If there is a provider who does not contribute to the coalition at all in the sense of (8) and is still being paid due to her potential for imaginary contribution assessed by the Shapley formula (1), which is not actually exploited in the current coalition, other players may improve their payoff sum by expelling the noncontributing provider

The condition $|Z| \geq 2$ plays an essential role in the theorem. For $|Z| \geq 2$, the concavity of the cost functions leads to the Shapley value not lying in the core, whereas, for the case $|Z| = 1$, the concavity of the cost function is proven to make the Shapley incentive structure lie in the core [4, Theorem 4.2].

### B. Convergence to the Grand Coalition

The notion of the core lends itself to the stability analysis of the grand coalition *on the assumption* that the players are already in the equilibrium, *i.e.*, the grand coalition. However, Theorem 2 still lets further questions unanswered. In particular, for the non-concave cost functions, it is unclear if the Shapley value is not in the core, which is still an open problem. We rather argue here that, whether the Shapley value lies in the core or not, the grand coalition is unlikely to occur by showing that the grand coalition is not a global attractor under some conditions.

To study the convergence of a game with coalition structure to the grand coalition, let us recall Definition 3. It is interesting that, though the notion of stability was not used in [4], one main argument of this work was that the system with one provider would converge to the grand coalition, hinting the importance of the following convergence result with multiple providers. The proof of the following theorem is given in Appendix D.

**Theorem 3 (A-D Payoff Does Not Lead to the Grand Coalition)** Suppose $|Z| \geq 2$ and $\widetilde{\Omega}_p(y)$ is not constant in the interval $y \in [0, x]$ for any $p \in Z$ where $x = |\bar{H}|/|H|$. The followings hold for all $p \in Z$ and $n \in \bar{H}$.
- The A-D payoff of provider $p$ in coalition $\{p\} \cup \bar{H}$ is larger than that in all coalition $T \cup \bar{H}$ for $\{p\} \subsetneq T \subseteq Z$.
- The A-D payoff of peer $n$ in coalition $\{p\} \cup \bar{H}$ is smaller than that in all coalition $T \cup \bar{H}$ for $\{p\} \subsetneq T \subseteq Z$.

In plain words, a provider, who is in cooperation with a peer set, will receive the highest dividend when she cooperates only with the peers excluding other providers whereas each peer wants to cooperate with as many as possible providers. It is surprising that, for the multiple provider case, *i.e.*, $|Z| \geq 2$, each provider benefits from forming a single-provider coalition *whether* the cost function is concave *or not*. There is no *positive* incentives for providers to cooperate with each other under the implementation of A-D payoffs. On the contrary, a peer always looses by leaving the grand coalition.

Upon the condition that each provider begins with a single-provider coalition with a sufficiently large number of peers, one cannot reach the grand coalition because some single-provider coalitions are already *stable* in the sense of the stability in Definition 3. That is, the grand coalition is not the global attractor. For instance, take $\mathcal{P} = \{\{p\} \cup H, \cdots\}$ as the current coalition structure where all peers are possessed by the provider $p$. Then it follows from Theorem 3 that players cannot make any transition from $\mathcal{P}$ to $\{\Phi \cup H, \cdots\}$ where $\Phi \subseteq Z$ is any superset of $\{p\}$ because provider $p$ will not agree to do so.

## V. A Critique of the A-D Payoff for Separate Providers

The discussion so far has focused on the stability of the grand coalition. The result in Theorem 2 suggests that if there is a noncontributing (free-riding) provider, which is true even for concave cost functions for multiple providers, the grand coalition will not be formed. The situation is aggravated by Theorem 3, stating that *single-provider coalitions* (*i.e.*, the separated case) will persist if providers are rational. We now illustrate the weak points of the A-D payoff under the single-provider coalitions with a few representative examples.

### A. Unfairness and Monopoly

**Example 1 (Unfairness)** Suppose that there are two providers, *i.e.*, $Z = \{p, q\}$, with $\widetilde{\Omega}_p(x) = 7(x-1)^{1.5}/8 + 1/8$ and $\widetilde{\Omega}_q(x) = 1 - x$, both of which are decreasing and *convex*. All values are shown in Fig. 3 as functions of $x$. In line with Theorem 3, providers are paid more than their Shapley values, whereas peers are paid less than theirs.

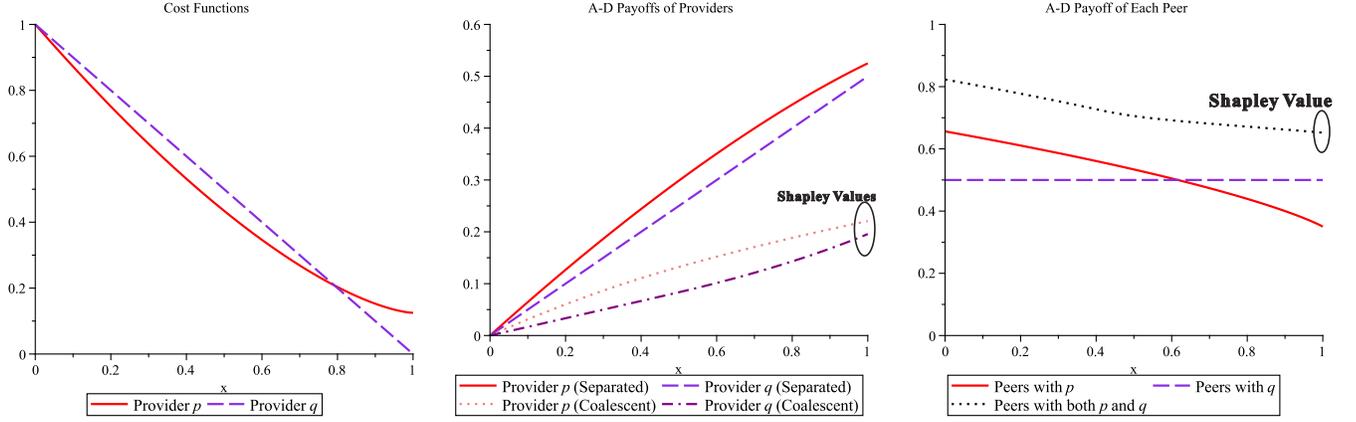

Fig. 3. Example 1: A-D Payoffs of Two Providers and Peers for Convex Cost Functions.

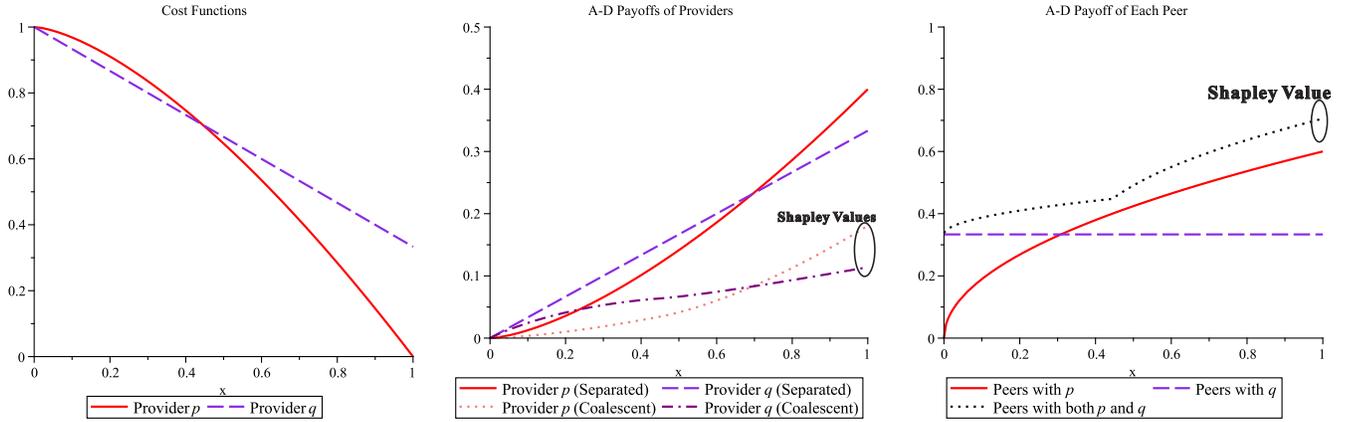

Fig. 4. Example 2: A-D Payoffs of Two Providers and Peers for Concave Cost Functions.

We can see that each peer $n$ will be paid $21/32$ ($\widetilde{\varphi}_n^{\{p\}}(0)$) when he is contained by the coalition $\{p, n\}$ and the payoff decreases with the number of peers in this coalition. On the other hand, provider $p$ wants to be assisted by as many peers as possible because $\widetilde{\varphi}_p^{\{p\}}(x)$ is increasing in $x$. If it is possible for $n$ to prevent other peers from joining the coalition, he can get $21/32$. However, it is more likely in real systems that no peer can kick out other peers, as discussed in [4, Section 5.1] as well. Thus, $p$ will be assisted by $x = 0.6163$ fraction of peers, which is the unique solution of $\widetilde{\varphi}_n^{\{p\}}(x) = \widetilde{\varphi}_n^{\{q\}}(x)$ while $q$ will be assisted by $1 - x = 0.3837$ fraction of peers.

**Example 2 (Monopoly)** Consider a two-provider system $Z = \{p, q\}$ with $\widetilde{\Omega}_p(x) = 1 - x^{3/2}$ and $\widetilde{\Omega}_q(x) = 1 - 2x/3$, both of which are decreasing and *concave*. All values including the Shapley values are shown in Fig. 4. Not to mention unfairness in line with Example 1 and Theorem 3, provider $p$ *monopolizes* the whole peer-assisted services. No provider has an incentive to cooperate with other provider. It can be seen that all peers will assist provider $p$ because $\widetilde{\varphi}_n^{\{p\}}(x) > \widetilde{\varphi}_n^{\{q\}}(x)$ for $x > 25/81$. Appealing to Definition 3, if the providers are initially separated, the coalition structure will converge to the service monopoly by $p$. In line with Lemma 2 and Theorem 2, even if the grand coalition is supposed to be the initial condition, it is not stable in the sense of the core. The noncontributing provider (Definition 5) in this example is $q$.

## B. Instability of A-D Payoff Mechanism

The last example illustrates the A-D payoff can even induce an analog of the limit cycle in nonlinear control theory.

**Example 3 (Oscillation)** Consider a game with two providers and two peers where $N = \{p_1, p_2, n_1, n_2\}$. If $\{n_1\}$, $\{n_2\}$ and $\{n_1, n_2\}$ assist the content distribution of $p_1$, the reduction of the distribution cost is respectively 10\$, 9\$ and 11\$ per month. However, the hard disk maintenance cost incurred from a peer is 5\$. In the meantime, if $\{n_1\}$, $\{n_2\}$ and $\{n_1, n_2\}$ assist the content distribution of $p_2$, the reduction of the distribution cost is respectively 6\$, 3\$ and 13\$ per month. In this case, the hard disk maintenance cost incurred from a peer is supposed to be 2\$ due to smaller contents of $p_2$ as opposed to those of $p_1$. We refer to Appendix E for a detailed analysis.

TABLE I
EXAMPLE 3: A-D PAYOFF AND BLOCKING COALITION $C$

|  | $\{p_1p_2, n_1n_2\}$ | $\{p_1p_2, n_1, n_2\}$ | $\{p_1, p_2, n_1n_2\}$ | $\{p_1, p_2, n_1, n_2\}$ | $\{p_1n_1, p_2n_2\}$ |
|---|---|---|---|---|---|
| $\varphi_{p_1}$ | 0 | 0 | 0 | 0 | 5/2=2.5 |
| $\varphi_{p_2}$ | 0 | 0 | 0 | 0 | 1/2=0.5 |
| $\varphi_{n_1}$ | 0 | 0 | 0 | 0 | 5/2=2.5 |
| $\varphi_{n_2}$ | 0 | 0 | 0 | 0 | 1/2=0.5 |
| $C$ | $p_1n_1, p_1n_2, p_2n_1, p_2n_2, p_1p_2n_1n_2, p_2n_1n_2$ | | | | $p_2n_1n_2$ |
| recurrent | X | X | X | X | O |

|  | $\{p_1p_2n_1n_2\}$ | $\{p_1p_2n_1, n_2\}$ | $\{p_2n_1n_2, p_1\}$ | $\{p_1p_2n_2, n_1\}$ | $\{p_1n_2, p_2, n_1\}$ |
|---|---|---|---|---|---|
| $\varphi_{p_1}$ | 7/6 = 1.17 | 7/6=1.17 | 0 | 5/3=1.67 | 2 |
| $\varphi_{p_2}$ | 19/6 = 3.17 | 2/3=0.67 | 23/6=3.83 | 1/6=0.17 | 0 |
| $\varphi_{n_1}$ | 17/6 = 2.83 | 19/6=3.17 | 10/3=3.33 | 0 | 0 |
| $\varphi_{n_2}$ | 11/6 = 1.83 | 0 | 11/6=1.83 | 13/6=2.17 | 2 |
| $C$ |  | $p_1n_2$ |  | $p_1n_1, p_2n_1$ |  |
| recurrent | X | X | O | X | X |

|  | $\{p_1n_1n_2, p_2\}$ | $\{p_1n_1, p_2, n_2\}$ | $\{p_1, n_1, p_2n_2\}$ | $\{p_1n_2, p_2n_1\}$ | $\{p_1, n_2, p_2n_1\}$ |
|---|---|---|---|---|---|
| $\varphi_{p_1}$ | 11/6=1.83 | 5/2=2.5 | 0 | 2 | 0 |
| $\varphi_{p_2}$ | 0 | 0 | 1/2=0.5 | 2 | 2 |
| $\varphi_{n_1}$ | -1/6=-0.17 | 5/2=2.5 | 0 | 2 | 2 |
| $\varphi_{n_2}$ | -2/3=-0.67 | 0 | 1/2=0.5 | 2 | 0 |
| $C$ | $p_1n_1, p_1n_2, p_2n_1, p_2n_2, p_2n_1n_2, n_1, n_2, n_1n_2$ | $p_2n_1n_2, p_2n_2$ | $p_1n_1, p_1n_2, p_2n_1, p_1p_2n_1n_2, p_2n_1n_2$ | $p_1n_1$ | $p_1n_1, p_1n_2, p_1p_2n_1n_2, p_2n_1n_2$ |
| recurrent | X | O | X | O | X |

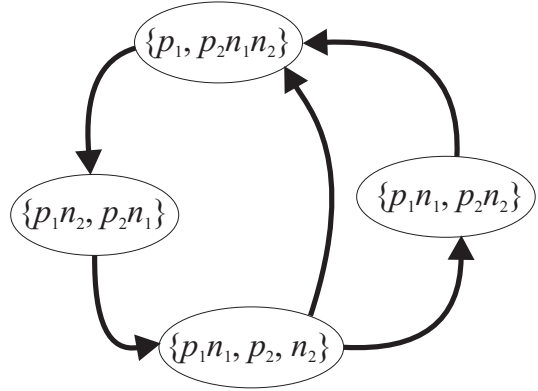

Fig. 5. Example 3: A-D Payoff Mechanism Leads to Oscillatory Coalition Structure.

Table I contains A-D payoffs (and Shapley payoffs for the grand coalition) and blocking coalitions $C \subseteq N$ for any coalition structure where, for notational simplicity, we adopt a simplified expression for coalitional structure $\mathcal{P}$: a coalition $\{a, b, c\} \in \mathcal{P}$ is denoted by $abc$ and each singleton set $\{i\}$ is denoted by $i$. We first observe that the Shapley payoff of this example does not lie in the core.

As time tends to infinity, the A-D payoff exhibits an oscillation of the partition $\mathcal{P}$ consisting of the four recurrent coalition structures as shown in Fig. 5. As of now, from the-state-of-the-art in the literature on this behavior [11], it is not yet clear how this behavior will be developed in large-scale systems.

## VI. CONCLUDING REMARKS

A quote from an interview of BBC iPlayer with CNET UK [14]: "*Some people didn't like their upload bandwidth being used. It was clearly a concern for us, and we want to make sure that everyone is happy, unequivocally, using iPlayer.*"

In this paper, we have studied whether the Shapley incentive structure in peer-assisted services would be in conflict with the pursuit of profits by rational content providers and peers. A lesson from our analysis is summarized as: even though it is righteous to pay peers more because they become relatively more useful as the number of peer-assisted services increases, the content providers will not admit that peers should receive their due deserts. The providers tend to persist in single-provider coalitions. In the sense of the classical stability notion, called 'core', the cooperation would have been broken even if we had begun with the grand coalition as the initial condition. Secondly, we have illustrated yet another problems when we use the Shapley-like incentive for the exclusive single-provider coalitions. These results suggest that the profit-sharing system, Shapley value, and hence its fairness axioms, are not compatible with the selfishness of the content providers.

## REFERENCES

[1] V. Valancius, N. Laoutaris, L. Massoulié, C. Diot, and P. Rodriguez, "Greening the Internet with Nano Data Centers," in *Proc. ACM CoNEXT*, Dec. 2009.
[2] M. Cha, P. Rodriguez, S. Moon, and J. Crowcroft, "On next-generation telco-managed P2P TV architectures," in *Proc. USENIX IPTPS*, Feb. 2008.
[3] RNCOS, "Global IPTV market forecast to 2014," *Market Research Report*, Feb. 2011.
[4] V. Misra, S. Ioannidis, A. Chaintreau, and L. Massoulié, "Incentivizing peer-assisted services: A fluid Shapley value approach," in *Proc. ACM Sigmetrics*, Jun. 2010.
[5] R. Aumann and L. Shapley, *Values of Non-Atomic Games*. Princeton University Press, 1974.
[6] S. Hart and M. Kurz, "Endogenous formation of coalitions," *Econometrica*, vol. 51, pp. 1047–1064, 1983.
[7] L. Shapley, *A Value for n-Person Games*. In H. W. Kuhn and A. W. Tucker, editors, Contribution to the Theory of Games II, vol. 28 of Annals of Mathematics Studies, Princeton University Press, 1953.
[8] R. Aumann and J. Drèze, "Cooperative games with coalition structures," *International Journal of Game Theory*, vol. 3, pp. 217–237, 1974.
[9] W. Saad, Z. Han, M. Debbah, A. Hjørungnes, and T. Başar, "Coalitional game theory for communication networks," *IEEE Signal Processing Mag.*, vol. 26, no. 5, pp. 77–97, 2009.
[10] B. Peleg and P. Sudhölter, *Introduction to the Theory of Cooperative Games*, 2nd ed. Springer-Verlag, 2007.
[11] A. Tutic, "The Aumann-Drèze value, the Wiese value, and stability: A note," *International Game Theory Review*, vol. 12, no. 2, pp. 189–195, 2010.
[12] E. Pinheiro, W. Weber, and L. A. Barroso, "Failure trends in a large disk drive population," in *Proc. USENIX FAST*, Feb. 2007.
[13] M. S. Bazaraa, H. D. Sherali, and C. M. Shetty, *Nonlinear Programming: Theory and Algorithms*, 2nd ed. John Wiley & Sons Inc., 1993.
[14] N. Lanxon, "iPlayer uncovered: What powers the BBC's epic creation?" *CNET UK*, May 2009.
[15] R. Myerson, "Graphs and cooperation in games," *Mathematics of Operstions Research*, vol. 2, pp. 225–229, 1977.

## APPENDIX

### A. An Intelligent Coalition for Non-Decreasing Worth Functions

It is reasonable to assume that, if the assistance from some peers *increases* the operational cost, they should not be allowed to assist the content distribution. In light of this, the worth of $S$ can be defined as a maximum of $\hat{v}(S)$:

$$v(S) := \Omega_p^\eta(0) - \min_{0 \leq y \leq x} \Omega_p^\eta(y). \tag{9}$$

There is a subtle difference between (2) and (9). Informally speaking, not all players within a coalition will remain passively when some outside peers are to decrease the worth of the coalition. Therefore, in games with worth function (2), peers who could have resulted in the decrease of the worth are not even eligible for membership of $S$. One can easily show that the core of game $(S, v)$, where $S \cap \bar{Z} = \{p\}$ and $|S \cap \bar{H}|/\eta = x$, is nonempty by using the Bondareva-Shapley Theorem [10, Theorem 3.1.4].

### B. Proof of Theorem 1

We use notation $\widetilde{\varphi}^{\bar{Z}}(x)$ to denote $\widetilde{\varphi}(\bar{Z} \cup \bar{H}, v)$. We use the mathematical induction to prove this theorem. The equation (5) holds for $|\bar{Z}| = 0$ and $\bar{Z} = \emptyset$ (empty set) because we have from (5) that there is no provider to pay and $\widetilde{\varphi}_n^\emptyset(x) = 0$ for $n \in \bar{H}$.

Now we **assume** that (5) holds for all $\Xi \subsetneq \bar{Z}$ such that $|\Xi| \leq \xi$ where $\xi \geq 0$. To prove Theorem 1, it *suffices* to show that (5) also holds for all $\Xi' \subseteq \bar{Z}$ such that $|\Xi'| = \xi + 1$. To this end, we first apply Axiom **CE**. As $\eta$ tends to infinity while $x$ remains unchanged, for $p \in \Xi'$ and $n \in \bar{H}$, Axiom **CE** for the partition $\{\Xi' \cup \bar{H}\}$ can be rewritten as follows:

$$\sum_{p \in \Xi'} \widetilde{\varphi}_p^{\Xi'}(x) + x \widetilde{\varphi}_n^{\Xi'}(x) = \sum_{p \in \Xi'} \widetilde{\Omega}_p(0) - M_\Omega^{\Xi'}(x) \tag{10}$$

which is the *normalized* (which we did in (4)) total coalition worth created by the coalition $\Xi' \cup \bar{H}$. Another axiom we apply is Axiom **FAIR** (fairness) which was used by Myerson [15] to characterize the Shapley value. It follows from **FAIR** that

$$\widetilde{\varphi}_n^{\Xi'}(x) - \widetilde{\varphi}_n^{\Xi' \setminus \{p\}}(x) = \frac{\mathrm{d}}{\mathrm{d}x} \widetilde{\varphi}_p^{\Xi'}(x), \quad \text{for all } p \in \Xi'. \tag{11}$$

Summing up these equation for $p \in \Xi'$ and dividing the sum by $|\Xi'| = \xi + 1$, we obtain

$$\widetilde{\varphi}_n^{\Xi'}(x) = \frac{1}{\xi+1} \sum_{p \in \Xi'} \left( \widetilde{\varphi}_n^{\Xi' \setminus \{p\}}(x) + \frac{\mathrm{d}}{\mathrm{d}x} \widetilde{\varphi}_p^{\Xi'}(x) \right) = \frac{1}{\xi+1} \sum_{p \in \Xi'} \widetilde{\varphi}_n^{\Xi' \setminus \{p\}}(x) + \frac{1}{\xi+1} \frac{\mathrm{d}}{\mathrm{d}x} \sum_{p \in \Xi'} \widetilde{\varphi}_p^{\Xi'}(x) \tag{12}$$

by plugging which into (10), we obtain

$$(\xi+1) \sum_{p \in \Xi'} \widetilde{\varphi}_p^{\Xi'}(x) + x \frac{\mathrm{d}}{\mathrm{d}x} \sum_{p \in \Xi'} \widetilde{\varphi}_p^{\Xi'}(x) = (\xi+1) \sum_{p \in \Xi'} \widetilde{\Omega}_p(0) - (\xi+1) M_\Omega^{\Xi'}(x) - x \sum_{p \in \Xi'} \widetilde{\varphi}_n^{\Xi' \setminus \{p\}}(x). \tag{13}$$

Since we know the form of $\widetilde{\varphi}_n^{\Xi' \setminus \{p\}}(x)$ for all $p \in \Xi'$ from the assumption ($\because |\Xi' \setminus \{p\}| = \xi$), (13) is an ordinary differential equation of unknown function $\sum_{p \in \Xi'} \widetilde{\varphi}_p^{\Xi'}(x)$. Denote the RHS of (13) by $G(x)$. Appealing to [4, Lemma 3], we get

$$\sum_{p \in \Xi'} \widetilde{\varphi}_p^{\Xi'}(x) = \int_0^1 u^\xi G(ux) du = \sum_{p \in \Xi'} \widetilde{\Omega}_p(0) - \int_0^1 u^\xi(\xi+1) M_\Omega^{\Xi'}(ux) du - \int_0^1 u^{\xi+1} x \sum_{p \in \Xi'} \widetilde{\varphi}_n^{\Xi' \setminus \{p\}}(ux) du,$$

$$\frac{d}{dx} \sum_{p \in \Xi'} \widetilde{\varphi}_p^{\Xi'}(x) = -(\xi+1) \int_0^1 u^{\xi+1} \frac{dM_\Omega^{\Xi'}}{dx}(ux) du - \int_0^1 u^{\xi+1} \sum_{p \in \Xi'} \widetilde{\varphi}_n^{\Xi' \setminus \{p\}}(ux) du - \int_0^1 u^{\xi+2} x \sum_{p \in \Xi'} \frac{d\widetilde{\varphi}_n^{\Xi' \setminus \{p\}}}{dx}(ux) du \quad (14)$$

$$= -(\xi+1) \int_0^1 u^{\xi+1} \frac{dM_\Omega^{\Xi'}}{dx}(ux) du + (\xi+1) \int_0^1 u^{\xi+1} \sum_{p \in \Xi'} \widetilde{\varphi}_n^{\Xi' \setminus \{p\}}(ux) du - \sum_{p \in \Xi'} \widetilde{\varphi}_n^{\Xi' \setminus \{p\}}. \quad (15)$$

where the last expression follows by integrating the last term of (14) by parts. From (12) and (15), $\widetilde{\varphi}_n^{\Xi'}(x)$ is rearranged as

$$\widetilde{\varphi}_n^{\Xi'}(x) = -\int_0^1 u^{\xi+1} \frac{dM_\Omega^{\Xi'}}{dx}(ux) du + \int_0^1 u^{\xi+1} \sum_{p \in \Xi'} \widetilde{\varphi}_n^{\Xi' \setminus \{p\}}(ux) du. \quad (16)$$

From the assumption, $\widetilde{\varphi}_n^{\Xi' \setminus \{p\}}(x)$ is given by (5) for $\bar{Z} = \Xi' \setminus \{p\}$, which is plugged into the last term of (16) to yield

$$\int_0^1 u^{\xi+1} \sum_{p \in \Xi'} \widetilde{\varphi}_n^{\Xi' \setminus \{p\}}(ux) du = -\sum_{p \in \Xi'} \sum_{S \subseteq \Xi' \setminus \{p\}} \int_0^1 \int_0^1 (ut)^{|S|} (u-ut)^{\xi-|S|} \frac{dM_\Omega^S}{dx}(utx) u dt du. \quad (17)$$

In the meantime, we need the following identity to reduce the double integral of (17):

$$\int_0^1 \int_0^1 (ut)^{|S|} (u-ut)^{\xi-|S|} f(utx) u dt du = \int_0^1 \int_0^u \tau^{|S|} (u-\tau)^{\xi-|S|} f(\tau x) d\tau du = \int_0^1 \int_\tau^1 \tau^{|S|} (u-\tau)^{\xi-|S|} f(\tau x) du d\tau$$

$$= \int_0^1 \frac{1}{\xi+1-|S|} \tau^{|S|} (1-\tau)^{\xi+1-|S|} f(\tau x) d\tau \quad (18)$$

where we used the change of variable $ut = \tau$ and changed the order of the double integration with respect to $u$ and $\tau$. Plugging (18) into (17) yields

$$\int_0^1 u^{\xi+1} \sum_{p \in \Xi'} \widetilde{\varphi}_n^{\Xi' \setminus \{p\}}(ux) du = -\sum_{p \in \Xi'} \sum_{S \subseteq \Xi' \setminus \{p\}} \frac{1}{\xi+1-|S|} \int_0^1 u^{|S|} (1-u)^{\xi+1-|S|} \frac{dM_\Omega^S}{dx}(ux) du$$

$$= -\sum_{S \subseteq \Xi', S \neq \Xi'} \int_0^1 u^{|S|} (1-u)^{\xi+1-|S|} \frac{dM_\Omega^S}{dx}(ux) du. \quad (19)$$

where the last equality holds because

$$\sum_{p \in \Xi'} \sum_{S \subseteq \Xi' \setminus \{p\}} f(S) : \sum_{S \subseteq \Xi', S \neq \Xi'} f(S) = (\xi+1) \cdot \binom{\xi}{|S|} : \binom{\xi+1}{|S|} = \xi + 1 - |S| : 1.$$

Plugging (19) into (16) establishes the following desired result:

$$\widetilde{\varphi}_n^{\Xi'}(x) = -\sum_{S \subseteq \Xi'} \int_0^1 u^{|S|} (1-u)^{\xi+1-|S|} \frac{dM_\Omega^S}{dx}(ux) du \quad (20)$$

from which follows

$$\widetilde{\varphi}_n^{\Xi'}(x) - \widetilde{\varphi}_n^{\Xi' \setminus \{p\}}(x) = -\sum_{S \subseteq \Xi'} \int_0^1 u^{|S|} (1-u)^{\xi+1-|S|} \frac{dM_\Omega^S}{dx}(ux) du + \sum_{S \subseteq \Xi' \setminus \{p\}} \int_0^1 u^{|S|} (1-u)^{\xi-|S|} \frac{dM_\Omega^S}{dx}(ux) du.$$

Since the first term of the RHS can be decomposed into the following:

$$-\sum_{S \subseteq \Xi' \setminus \{p\}} \int_0^1 u^{|S|+1} (1-u)^{\xi+1-(|S|+1)} \frac{dM_\Omega^{S \cup \{p\}}}{dx}(ux) du - \sum_{S \subseteq \Xi' \setminus \{p\}} \int_0^1 u^{|S|} (1-u)^{\xi+1-|S|} \frac{dM_\Omega^S}{dx}(ux) du,$$

we can obtain

$$\widetilde{\varphi}_n^{\Xi'}(x) - \widetilde{\varphi}_n^{\Xi' \setminus \{p\}}(x) = -\sum_{S \subseteq \Xi' \setminus \{p\}} \int_0^1 u^{|S|+1} (1-u)^{\xi-|S|} \left( \frac{dM_\Omega^{S \cup \{p\}}}{dx}(ux) - \frac{dM_\Omega^S}{dx}(ux) \right) du. \quad (21)$$

Integrating (11) from 0 to $x$ with respect to $x$ and from (21), we get

$$\widetilde{\varphi}_p^{\Xi'}(x) = -\sum_{S \subseteq \Xi' \setminus \{p\}} \int_0^1 u^{|S|}(1-u)^{\xi-|S|} \left( M_\Omega^{S \cup \{p\}}(ux) - M_\Omega^S(ux) - \underbrace{\left( M_\Omega^{S \cup \{p\}}(0) - M_\Omega^S(0) \right)}_{\widetilde{\Omega}_p(0)} \right) du$$

$$= \widetilde{\Omega}_p(0) - \sum_{S \subseteq \Xi' \setminus \{p\}} \int_0^1 u^{|S|}(1-u)^{\xi-|S|} \left( M_\Omega^{S \cup \{p\}}(ux) - M_\Omega^S(ux) \right) du$$

which finally establishes that (5) also holds for all $\Xi' \subseteq \bar{Z}$ such that $|\Xi'| = \xi + 1$, hence completing the proof.

### C. Proof of Theorem 2

To prove the theorem, we need to show that the condition for the core in Definition 1 is violated, implying that it suffices to show the following:

$$\widetilde{\varphi}_p^Z(1) > \sum_{i \in Z} \widetilde{\Omega}_i(0) - M_\Omega^Z(1) - \left( \sum_{i \in Z \setminus \{p\}} \widetilde{\Omega}_i(0) - M_\Omega^{Z \setminus \{p\}}(1) \right) = \widetilde{\Omega}_p(0) - \left( M_\Omega^Z(1) - M_\Omega^{Z \setminus \{p\}}(1) \right). \quad (22)$$

This means that the payoff of $p \in Z$ is greater than the marginal increase of the limit worth, *i.e.*,

$$\lim_{\eta \to \infty} \frac{1}{\eta} v(Z \cup H) - \lim_{\eta \to \infty} \frac{1}{\eta} v((Z \setminus \{p\}) \cup H).$$

Subtracting the RHS of (22) from the LHS of (22) and using the expression of $\widetilde{\varphi}_p^Z(1)$ in (5), we have

$$M_\Omega^Z(1) - M_\Omega^{Z \setminus \{p\}}(1) - \sum_{S \subseteq Z \setminus \{p\}} \int_0^1 u^{|S|}(1-u)^{|Z|-1-|S|} \left( M_\Omega^{S \cup \{p\}}(u) - M_\Omega^S(u) \right) du. \quad (23)$$

We can see from Definition 5 that $M_\Omega^Z(1) - M_\Omega^{Z \setminus \{p\}}(1) = \widetilde{\Omega}_p(0)$. From the assumption, there exists a noncontributing provider which we denote by $p$. To show that (23) is strictly positive, we rewrite the last factor of the integrand as follows:

$$M_\Omega^{S \cup \{p\}}(y) - M_\Omega^S(y) = \min \left\{ \sum_{i \in S \cup \{p\}} \widetilde{\Omega}_i(y_i) \mid \sum_{i \in S \cup \{p\}} y_i \leq y, y_i \geq 0 \right\} - \min \left\{ \sum_{i \in S} \widetilde{\Omega}_i(y_i) \mid \sum_{i \in S} y_i \leq y, y_i \geq 0 \right\}$$

where the first term in the RHS can be rearranged as

$$\min \left\{ \sum_{i \in S \cup \{p\}} \widetilde{\Omega}_i(y_i) \mid \sum_{i \in S \cup \{p\}} y_i \leq y, y_i \geq 0 \right\} \leq \widetilde{\Omega}_p(0) + \min \left\{ \sum_{i \in S} \widetilde{\Omega}_i(y_i) \mid \sum_{i \in S} y_i \leq y, y_i \geq 0 \right\}$$

where the inequality holds from that $\widetilde{\Omega}_i(y)$, $i \in Z$, are non-increasing. It can be easily seen that the inequality holds by considering two cases $y_p = 0$ and $y_p > 0$. The inequality becomes *strict* when $S = \emptyset$ over some interval in $[0, x]$ whose length is positive due to the assumption that $\widetilde{\Omega}_p(y)$ is not constant in the interval $y \in [0, x]$ and non-increasing. From this, we can see that (23) is greater than

$$\widetilde{\Omega}_p(0) - \sum_{S \subseteq Z \setminus \{p\}} \int_0^1 u^{|S|}(1-u)^{|Z|-1-|S|} \widetilde{\Omega}_p(0) du = 0$$

which establishes (22), hence completing the proof.

### D. Proof of Theorem 3

To prove Theorem 3, it suffices to show that the following holds for $\{p\} \subsetneq T$ such that $T \subseteq Z$:

$$\varphi_p^{\{p\}}(x) - \varphi_p^T(x) = \sum_{S \subseteq T \setminus \{p\}} \int_0^1 u^{|S|}(1-u)^{|T|-1-|S|} \left( M_\Omega^{S \cup \{p\}}(ux) - M_\Omega^S(ux) \right) du - \int_0^1 M_\Omega^{\{p\}}(ux) du > 0 \quad (24)$$

which implies that the payoff of $p$ when it is the only provider of the coalition is larger than that with other providers $T \setminus \{p\}$. To this end, we first observe that, for $y \leq x$,

$$M_\Omega^{S \cup \{p\}}(y) - M_\Omega^S(y) = \min\left\{\sum_{i \in S \cup \{p\}} \widetilde{\Omega}_i(y_i) \mid \sum_{i \in S \cup \{p\}} y_i \leq y, y_i \geq 0\right\} - \min\left\{\sum_{i \in S} \widetilde{\Omega}_i(y_i) \mid \sum_{i \in S} y_i \leq y, y_i \geq 0\right\}.$$

Here the first term in the RHS can be rearranged as

$$\min\left\{\sum_{i \in S \cup \{p\}} \widetilde{\Omega}_i(y_i) \mid \sum_{i \in S \cup \{p\}} y_i \leq y, y_i \geq 0\right\} \geq M_\Omega^{\{p\}}(y) + \min\left\{\sum_{i \in S} \widetilde{\Omega}_i(y_i) \mid \sum_{i \in S} y_i \leq y, y_i \geq 0\right\}$$

where the inequality holds from that $M_\Omega^{\{i\}}(y)$, $i \in T$, are non-increasing. It can be easily seen that the inequality holds by considering two cases $y_p = 0$ and $y_p > 0$. The inequality becomes *strict* when $S = \emptyset$ over some interval in $[0, x]$ whose length is positive due to the assumption that $\widetilde{\Omega}_p(y)$ is not constant in the interval $y \in [0, x]$ and non-increasing. From this inequality, we have $M_\Omega^{S \cup \{p\}}(y) - M_\Omega^S(y) \geq M_\Omega^{\{p\}}(y)$ and the inequality is strict over some interval of positive length. Plugging this relation into (24) yields

$$\varphi_p^{\{p\}}(x) - \varphi_p^T(x) = \left(\sum_{S \subseteq T \setminus \{p\}} \int_0^1 u^{|S|} (1-u)^{|T|-1-|S|} M_\Omega^{\{p\}}(ux) du\right) - \int_0^1 M_\Omega^{\{p\}}(ux) du > 0.$$

In the meantime, we can see the following from (10):

$$\lim_{\eta \to \infty} v(\{p\} \cup \bar{H})/\eta = \widetilde{\Omega}_p(0) - M_\Omega^{\{p\}}(x) \leq \sum_{i \in T} \widetilde{\Omega}_i(0) - M_\Omega^T(x) = \lim_{\eta \to \infty} v(T \cup \bar{H})/\eta$$

which, when combined with $\varphi_p^{\{p\}}(x) > \varphi_p^T(x)$, implies the second part of the theorem.

*E. Computation of the A-D Payoff in Example 3*

From the description of the cost reduction and the hard disk maintenance cost incurred from peers, we can compute the *net* cost reduction for all possible coalitions. For example, if $n_1$ and $n_2$ help $p_1$, the coalition worth becomes $v(\{n_1, n_2, p_1\}) = 11\$ - 5\$ - 5\$ = 1\$$. In a similar way, we have the following result:

$$\hat{v}(S) = \begin{cases} 0, & \text{if } S \text{ is not profitable,} \\ 5, & \text{if } S = \{p_1, n_1\}, \\ 4, & \text{if } S = \{p_1, n_2\}, \\ 1, & \text{if } S = \{p_1, n_1, n_2\}, \\ 4, & \text{if } S = \{p_2, n_1\}, \\ 1, & \text{if } S = \{p_2, n_2\}, \\ 9, & \text{if } S = \{p_2, n_1, n_2\}. \end{cases}$$

Using the same expression (3) as in Section III, it is easy to see that the coalition worths for coalescent provider cases are $v(\{p_1, p_2, n_1\}) = 5$, $v(\{p_1, p_2, n_2\}) = 4$ and $v(\{p_1, p_2, n_1, n_2\}) = 9$.

Suppose that peers continue to form a new coalition $C'$ when they can improve away from the current coalition $C$. That is, if there is a blocking coalition $C'$ (Definition 3), they will betray $C$. It is easy to see that almost all coalition structures are *transient*. Note that, as discussed in [6], one can get two types of resulting coalition structures when a player departs from a coalition. For example, if player $n_1$ departs from her coalition in $\{p_1 n_1 n_2, p_2\}$ to form a coalition with $p_2$, we may get either $\{p_1, n_2, p_2 n_1\}$ or $\{p_1 n_2, p_2 n_1\}$. To obtain Table I, we assumed only the latter case to simplify the exposition.